\documentclass{emulateapj}

\usepackage{graphicx}
\usepackage{float}
\usepackage{amsmath}
\usepackage{epsfig,floatflt}

\begin{document}

\title{Comment on ``CCC-predicted low-variance circles in CMB sky and LCDM''}

\author{H.\ K.\ Eriksen\altaffilmark{1,2} and I. K. Wehus\altaffilmark{3,4}}

\email{h.k.k.eriksen@astro.uio.no}
\email{i.k.wehus@fys.uio.no}

\altaffiltext{1}{Institute of Theoretical Astrophysics, University of
  Oslo, P.O.\ Box 1029 Blindern, N-0315 Oslo, Norway}

\altaffiltext{2}{Centre of Mathematics for Applications, University of
  Oslo, P.O.\ Box 1053 Blindern, N-0316 Oslo, Norway}

\altaffiltext{3}{Theoretical Physics, Imperial College London, London
  SW7 2AZ, UK}

\altaffiltext{4}{Department of Physics, University of
  Oslo, P.O.\ Box 1048 Blindern, N-0316 Oslo, Norway}


\begin{abstract}
In a recent preprint ``CCC-predicted low-variance circles in the CMB
sky and LCDM'', \citet{gurzadyan:2011a} claim for the second time to
find evidence for pre-Big Bang activity in the form of concentric
circles of low variance in the WMAP data. The same claim was made in
November 2010, but quickly shown to be false by three independent
groups. The culprit was simply that Gurzadyan and Penrose's
simulations were based on an inappropriate power spectrum. In the most
recent paper, they now claim that the significance is indeed low if
the simulations are based on the realization-specific WMAP spectrum
(ie., the one directly measured from the sky maps and affected by
cosmic variance), but not if the simulations are based on a
\emph{theoretical} $\Lambda$CDM spectrum. In this respect, we note
that the three independent reanalyses all based their simulations on
the $\Lambda$CDM spectrum, not the observed WMAP spectrum, and this
alone should suffice to show that the updated claims are also
incorrect. In fact, it is evident from the plots shown in their new
paper that the spectrum is still incorrect, although in a different
way than in their first paper. Thus, Gurzadyan and Penrose's new
claims are just as wrong as those made in the first paper, and for the
same reason: The simulations are not based on an appropriate power
spectrum. Still, while this story is of little physical interest, it
may have some important implications in terms of scienctific
sociology: Looking back at the background papers leading up to the
present series by Gurzadyan and Penrose, in particular one introducing
the Kolmogorov statistic, we believe one can find evidence that a
community based and open access referee process may be more efficient
at rejecting incorrect results and claims than a traditional journal
based approach.
\end{abstract}

\section{The low-variance ring claims}
\label{sec:introduction}

In November 2010, \citet{gurzadyan:2010} posted a paper on the arXiv
preprint server claiming to find evidence of pre-Big Bang activity in
the cosmic microwave background temperature fluctuations as measured
by WMAP. These signatures were defined in terms of concentric rings of
``low variance'', presumably the result of violent collisions between
super-massive black holes and corresponding shock waves. The
statistical significance of these detections were reported to be more
than $6\sigma$. If true, this would indeed be a spectacular result.

However, three independent analyses by \citet{wehus:2011},
\citet{moss:2011} and \citet{hajian:2010} (the two former published on
the very same day) quickly showed that the results were flawed. The
problem was simply that Gurzadyan and Penrose had based their
simulations on an inappropriate power spectrum, effectively assuming
that the CMB consists of uncorrelated white noise in pixel space.

Very recently, in a paper called ``CCC-predicted low-variance circles
in the CMB sky and LCDM'', \citet{gurzadyan:2011a} make a second
attempt at the same claim, this time aiming to build their simulations
with a proper power spectrum. Specifically, they claim that if the
random simulation is built from the ``observed WMAP spectrum'', ie.,
the realization specific spectrum as directly measured by WMAP, the
statistical signficance of the rings is low, in agreement with the
results of the three independent reanalyses. However, if the
simulations are instead based on a theoretical (smooth) $\Lambda$CDM
spectrum, they claim that the rings are significant. Their hypothesis
is thus that the bumps and wiggles in the WMAP spectrum from cosmic
variance carries extra information, leading to a greater probability
of generating coherent rings (or vice versa, depending on ones
point-of-view).

Here it is worth noting a few facts. First, as clearly stated in each
of the three reanalysis papers, the simulations used in each case were
in fact based on the best-fit $\Lambda$CDM spectrum, not the
realization-specific WMAP spectrum. This is in accordance with the
generally accepted procedure for generating random CMB simulation;
usage of constrained spectrum realization is a very special case, and
would clearly warrant special justification. This point alone shows
that the updated claims by Gurzadyan and Penrose are still wrong: One
does of course find similar rings with a $\Lambda$CDM spectrum, and
not only with the WMAP spectrum. (This is rather obvious, as the two
spectra are by construction very similar, and the two-point correlation
function, which is really what enters into these calculations, is also
correspondingly similar.)

Second, we note that one can see directly from the figures shown in
Gurzadyan and Penrose's paper that their claimed ``$\Lambda$CDM''
spectrum is still flawed:
\begin{itemize}
\item Comparing Figures 1a and c, showing the mean variance profile,
  we see (as also Gurzadyan and Penrose note) that their $\Lambda$CDM
  spectrum leads to a significantly lower variance than in the
  observed data. Since this curve represents an \emph{average}
  variance quantity, it is given fully by the power spectrum, and it
  must therefore (up to cosmic variance) have the same mean as the
  observed WMAP realization. The fact that it is lower implies that
  the spectrum used for generating the simulations is also lower.
\item Gurzadyan and Penrose also note that the observed WMAP variance
  profile drops with decreasing radius, while the simulated profile
  does not. This feature was also noted in our original reanalysis
  paper \citep{wehus:2011}: This function \emph{has} to behave like
  this, simply because the CMB field is correlated and smoothed with
  an instrumental beam. On very small scales, the CMB variance
  converges to zero due to smoothing. The fact that the simulated
  example does not fall, therefore implies that it is wrong.
\item Third, Gurzadyan and Penrose also note that the simulated map in
  their Figure 5 is notably ``greener'' than the observed map in
  Figure 4, corresponding to having values closer to the average. This
  statement is fully equivalent to saying that the simulation has less
  power than the observed data -- and therefore that the simulation is
  wrong. 
\end{itemize}

It is difficult to say exactly what went wrong in the generation of
these updated simulations. At least they have a non-flat power
spectrum, which is a clear improvement over the first version. Still,
it is also clear that the present claims are still not correct, and
the same criticisms that were presented by \citet{wehus:2011},
\citet{moss:2011} and \citet{hajian:2010} still apply: When making
claims similar to those of \citet{gurzadyan:2011a}, it is
\emph{essential} to construct the underlying simulations with absolute
data fidelity.

\section{The Kolmogorov statistic and the refereeing process}
\label{sec:kolmogorov}

While the physical importance of Gurzadyan and Penrose's recent claims
in our opinion are marginal at best, we do believe that there are some
interesting points in terms of science sociology and the currently
accepted refereeing process. While performing the first reanalysis of
Gurzadyan and Penrose's claims, we read through most of the papers
cited in their original paper, trying to understand the background for
their claims.

In particular, one apparently central line of reasoning of
\citet{gurzadyan:2010} was based on the notion of the ``Kolmogorov
statistic'', as introduced by \citet{gurzadyan:2008, gurzadyan:2009}
and references therein. This statistic measures the degree of
``randomness'' within a set of stochastic variables. In particular,
\citet{gurzadyan:2011b} applied this statistic to the small disks in
the WMAP sky maps, and measured the degree of randomness within each
disk. The main conclusion drawn from this work was that only 20\% of
the signal was ``random'', while 80\% of the signal was
``non-random''. 

When reading these papers, it seems clear to us that Gurzadyan et
al.\ confuse randomness with correlation: While the CMB field is (most
likely) a random field, it is \emph{not} uncorrelated. Instead, the
CMB field is a smooth field on scales comparable with the instrumental
beam, and it has a well-defined non-flat power spectrum. Thus, the
real-space correlations are strong. Of course, the instrumental
\emph{noise} is virtually uncorrelated, and so there are indeed two
components here, one correlated and one uncorrelated. But neither is
non-random.

The interesting part of this story, though, is the fact that at least
five papers on this very topic have been accepted and published by the
reputable (and refereed) journal ``Astronomy and Astrophysics''. One
of these papers (called ``A weakly random universe?'') was even
published as a Letter, with an abstract stating that ``Deriving the
empirical Kolmogorov's function in the Wilkinson Microwave Anisotropy
Probe's maps, we obtain the fraction of the random signal to be about
20 per cent, i.e. the cosmological sky is a weakly random one.'' These
are truly extraordinary claims, and in our view have no root in
reality. Further, these claims are not irrelevant, as clearly
demonstrated by the most recent developments concerning the concentric
rings: They have, at the very least implicitly, led to an excessive
amount of publicity in the general public, potentially damaging the
public perception of cosmologists in a wider sense. In our view, this
is a clear demonstration of the potential weaknesses of the
established refereeing processes: Marginal, or even plain wrong, work
can be published due to an unattentive referee.

Contrary to this, it is interesting to note the reaction that came
after the first Gurzadyan and Penrose paper was put on the arXiv in
November 2011: In only a matter of weeks, three independent groups
refuted the original claim. Of course, this reaction was largely
triggered by the massive media attention that the original story got,
and which most papers will never experience. Nevertheless, we believe
that this particular case is a good demonstration of the power of
community review outside the established journals: The open community
can be a far more efficient reviewer than a somewhat arbitrary referee
appointed by a given journal.

\end{document}